\pdfoutput=1

\documentclass[]{spie}  %>>> use for US letter paper
\usepackage{amsmath,amsfonts,amssymb}
\usepackage{graphicx}
\usepackage[colorlinks=true, allcolors=blue]{hyperref}
\usepackage{here}

\title{Current Status and Future Plan of Osaka Prefecture University 1.85-m mm-submm Telescope Project}

\author[a]{Atsushi Nishimura}
\author[a]{Kazuki Tokuda}
\author[a]{Ryohei Harada}
\author[a]{Yutaka Hasegawa}
\author[a]{Shota Ueda}
\author[a]{Sho Masui}
\author[a]{Ryotaro Konishi}
\author[a]{Yasumasa Yamasaki}
\author[a]{Hiroshi Kondo}
\author[a]{Koki Yokoyama}
\author[a]{Takeru Matsumoto}
\author[a]{Taisei Minami}
\author[a]{Masanari Okawa}
\author[a]{Shinji Fujita}
\author[a]{Ayu Konishi}
\author[a]{Yuka Nakao}
\author[a]{Shimpei Nishimoto}
\author[a]{Sana Kawashita}
\author[a]{Sho Yoneyama}
\author[a]{Tatsuyuki Takashima}
\author[a]{Kenta Goto}
\author[a]{Nozomi Okada}
\author[a]{Kimihiro Kimura}
\author[a]{Yasuhiro Abe}
\author[a]{Kazuyuki Muraoka}
\author[a]{Hiroyuki Maezawa}
\author[a]{Toshikazu Onishi}
\author[a]{Hideo Ogawa}

\affil[a]{Department of Physical Science, Graduate School of Science, Osaka Prefecture University, 1-1 Gakuen-cho, Naka-ku, Sakai, Osaka 599-8531, Japan}

\authorinfo{Further author information: (Send correspondence to A.N.)\\A.N.: E-mail: nishimura@p.s.osakafu-u.ac.jp}

\begin{document}
\maketitle

\begin{abstract}
We report the current status of the 1.85-m mm-submm telescope installed at the Nobeyama Radio Observatory (altitude 1400 m) and the future plan. 
The scientific goal is to reveal the physical/chemical properties of molecular clouds in the Galaxy by obtaining large-scale distributions of molecular gas with an angular resolution of several arcminutes. 
A semi-automatic observation system created mainly in Python on Linux-PCs enables effective operations. 
A large-scale CO $J=$2--1 survey of the molecular clouds (e.g., Orion-A/B, Cygnus-X/OB7, Taurus-California-Perseus complex, and Galactic Plane), and a pilot survey of emission lines from minor molecular species toward Orion clouds have been conducted so far. 
The telescope also is providing the opportunities for technical demonstrations of new devices and ideas. 
For example, the practical realizations of PLM (Path Length Modulator) and waveguide-based sideband separating filter, installation of the newly designed waveguide-based circular polarizer and OMT (Orthomode Transducer), and so on. 
As the next step, we are now planning to relocate the telescope to San Pedro de Atacama in Chile (altitude 2500 m), and are developing very wideband receiver covering 210--375 GHz (corresponding to Bands 6--7 of ALMA) and full-automatic observation system. 
The new telescope system will provide large-scale data in the spatial and frequency domain of molecular clouds of Galactic plane and Large/Small Magellanic Clouds at the southern hemisphere. 
The data will be precious for the comparison with those of extra-galactic ones that will be obtained with ALMA as the Bands 6/7 are the most efficient frequency bands for the surveys in extra-galaxies for ALMA.
\end{abstract}


\section{INTRODUCTION}
\label{sec:intro}

The rotational transition line of CO is one of the most important emission line in astrophysics to investigate the distribution, physical properties, and kinematics of the interstellar medium (ISM).
Especially, the lines of CO $J=$1--0 and its isotopes have been widely used as a probe of the mass of molecular contents of the ISM, because of its high abundance, low critical density for excitation, and relatively stable chemical properties.
The distribution of the line in the Galaxy has been explored with relatively small-aperture telescopes, such as the 1.2-m telescopes at the Harvard-Smithsonian Center for Astrophysics (CfA) and Cerro Tololo Inter-American Observatory in Chile\cite{1985ApJ...297..751D, 1987ApJ...322..706D, 2001ApJ...547..792D}, as well as the 4-m telescope in Nagoya and the NANTEN telescope in Chile\cite{1998IAUS..179..165F, 2004dimg.conf..203O, 2004ASPC..317...59M, 2020SPIE.nishimura.nasco}.

Although the $J=$1--0 lines of CO are powerful tools to investigate the mass of the molecular cloud, the other transitions with different critical densities for the excitation are needed to investigate the local density and the temperature of the ISM.
However, such observations targeting the higher transition lines of CO were conduced only at coarse angular resolutions such as a survey of 60-cm telescope\cite{1994ApJ...425..641S, 1995ApJS..100..125S, 2001ApJ...551..794S, 2010PASJ...62.1277Y} or only toward small regions such as a survey of JCMT/HARP\cite{2010MNRAS.401..204B, 2012MNRAS.422..521B}, at the time of early 2010s, mainly because the development of sensitive receivers at high frequencies has been very difficult and also because the opacity of the Earth's atmosphere is high at low altitude site.

The 1.85-m mm-submm telescope was developed to realize a large-scale CO $J=$2--1 survey.
The first science operation was started on November 2011 by using a sideband-separating (2SB) SIS receiver, and the CO survey was continued several years.
The developments of receiver was continued during the operation, and the 1.85-m telescope was also used as a test bench of the new receiver concepts.
In this paper, we describe the history of the developments and observations of the 1.85-m telescope, current status of the project and the future plan in detail.

\section{DEVELOPMENT OF RADIO INSTRUMENTS}
\label{sec:inst}

\begin{figure} [ht]
\begin{center}
\begin{tabular}{c} %% tabular useful for creating an array of images 
\includegraphics[width=9cm, bb=0 0 480 410]{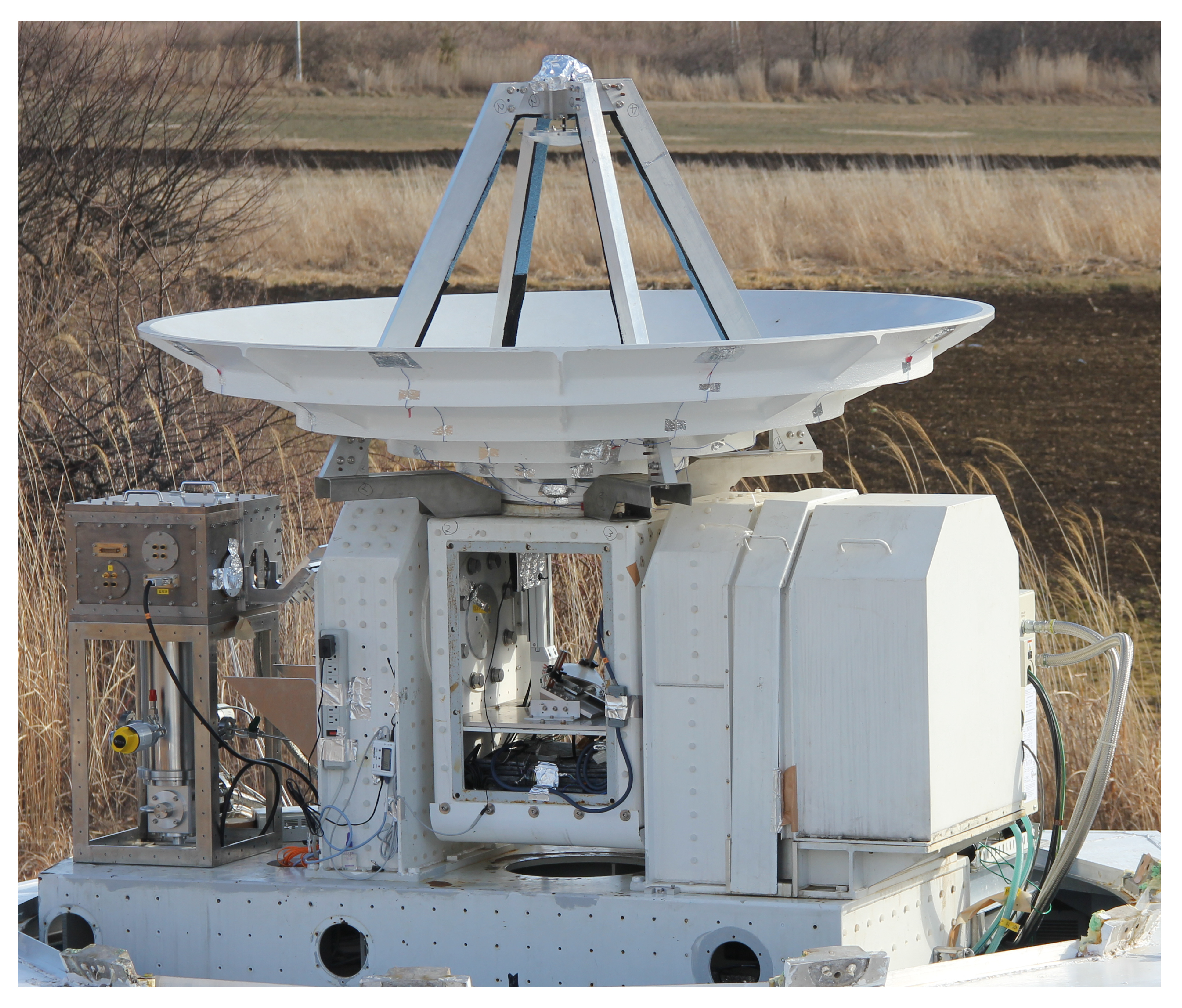}
\end{tabular}
\end{center}
\caption[example] 
%>>>> use \label inside caption to get Fig. number with \ref{}
{ \label{fig:telescope} 
Photograph of the 1.85-m telescope at the Nobeyama Radio Observatory.
}
\end{figure}

The 1.85-m telescope\cite{2013PASJ...65...78O} has a Cassegrain reflector antenna with Nasmyth beam-waveguide feed, and is mounted on an azimuth-elevation rotating structure (Fig \ref{fig:telescope}).
The main reflector is made of carving of one piese of aluminium, and its surface accuracy was measured to be 19 $\mu$m by a 3-D coordinate measuring machine, SNK MM-3500.
The radome is equipped to protect the telescope from wind, precipitation, and sunlight, that was originally used for the 1.2-m telescope installed at the summit of Mt. Fuji\cite{2000SPIE.4015..185S}.
The path length modulator\cite{2002.JIMRW.Mizuno} (PLM) was installed in front of the RF window of the receiver to reduce standing waves between the RF window and the calibration unit or the Al frames of the radome.
For more detail on the antenna and optics, see Ref. \citenum{2013PASJ...65...78O}.

\begin{figure} [ht]
\begin{center}
\begin{tabular}{c} %% tabular useful for creating an array of images 
\includegraphics[width=10cm, bb=0 0 956 508]{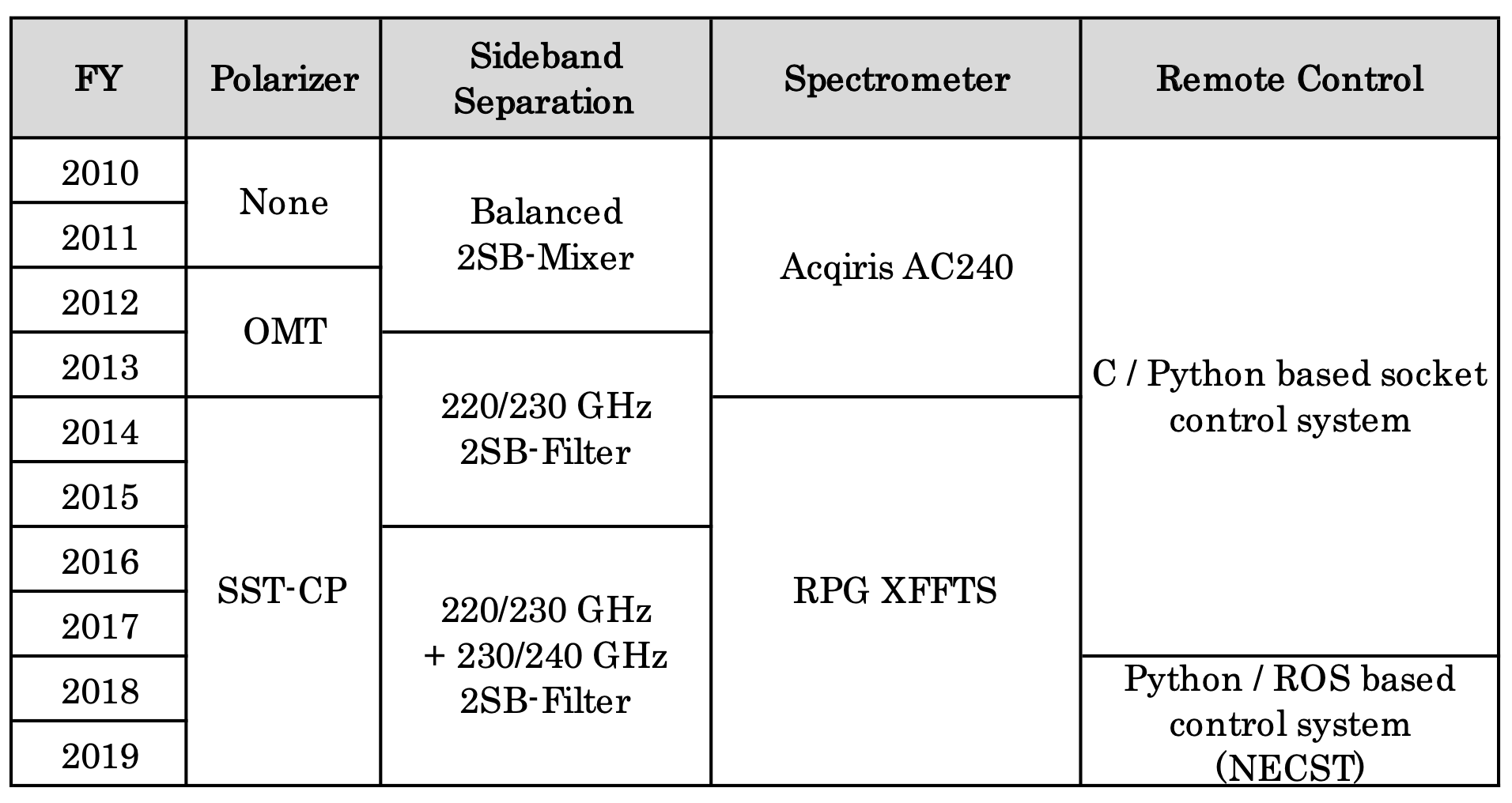}
\end{tabular}
\end{center}
\caption[example] 
%>>>> use \label inside caption to get Fig. number with \ref{}
{ \label{fig:history} 
History of the installed receiver and control system of the 1.85-m telescope.
}
\end{figure}

The construction of the telescope at the Nobeyama Radio Observatory (NRO) was started on 2007, and the first simultaneous observations of the $^{12}$CO, $^{13}$CO, and C$^{18}$O of $J=$2--1 transition lines were achieved in September 2009.
The beam size was measured to be 2.7$'$ at 230 GHz band.
Fig. \ref{fig:history} summarizes the receiver system and control system developed and used for the 1.85-m telescope.

\subsection{Receiver System}
\label{sec:rx}
\subsubsection{Season 2009--2012}
The first light receiver for the 1.85-m telescope was a balanced-type sideband-separating (2SB) SIS receiver.
The basic design is the same as Ref. \citenum{2007PASJ...59.1005N} which was developed for the 60-cm AMANOGAWA telescope.
The target LO frequency is $\sim225$ GHz, and the receiver outputs two IF signals with a frequency of 4--8 GHz.
In order to achieve simultaneous observations of $^{12}$CO $J=$2--1 (230.538 GHz), $^{13}$CO $J=$2--1 (220.399 GHz), and C$^{18}$O $J=$2--1 (219.560 GHz) by using one spectrometer with a bandwidth of 1 GHz, Acqiris AC240\cite{2005A&A...442..767B}, two output signals of the SIS receiver (1st IF) were split to 3 signals, then each of the signal was further down-converted to different frequency in the 2nd IF band, and finally those 3 signals merged and one 0--1 GHz signal which contains the 3 different RF bands was generated in the IF system\cite{2013PASJ...65...78O}.

In the observation season of 2012, an Orthomode Transducer\cite{2009.JIMRW.Asayama, 2010.JIMRW.Kamikura} (OMT) for dual-polarization observations was developed and installed to the telescope.
The measured characteristics of the OMT were an insertion loss of less than 0.5 dB, return loss of better than 20 dB, and cross polarization coupling of better than 20 dB across 210--250 GHz\cite{2013PASJ...65...78O}. 
Using the OMT with two balanced-type 2SB SIS receivers, observation efficiency was improved by a factor of 2 than previous seasons.

\subsubsection{Season 2013--2015}

Although the balanced-type 2SB SIS mixer receivers are very powerful to observe wide frequency ranges simultaneously, its image rejection ratio (IRR) is relatively low\cite{2009.JIMRW.Nakajima, 2016SPIE.9914E..1ZM}, typically $10$ dB, and the stability of the IRR is not enough especially for the long-term observations\cite{2016EP&S...68...34O}.
The 2SB-Filter\cite{2015JIMTW..36..445A} (2SBF) receiver has a high IRR typically more than $20$ dB, and the long-term stability.
In the 2SBF receiver, the cutout frequency of the lower sideband (LSB) and upper sideband (USB) are determined at the design of the filter implemented in waveguide.
The target emission lines of the 1.85-m telescope was CO $J=$2--1, and other lines were not planned to observe the season, thus it was not a problem even if the LO frequency was fixed.
We developed a 2SBF receiver optimized for $^{12}$CO, $^{13}$CO, and C$^{18}$O $J=$2--1 lines observations\cite{2017PASJ...69...91H}.
The IRR of the receiver was measured to be better than $25$ dB.

The spectrometer was replaced on the season of 2013 with a bandwidth of 2 GHz model, RPG XFFTS\cite{2012A&A...542L...3K}.
We installed four XFFTS boards, and use each board for each receiver output: one board for the $^{12}$CO band and another board for the $^{13}$CO and C$^{18}$O band.

In the observation season of 2014, we replaced the OMT to a stepped septum-type waveguide circular polarizer\cite{2017.JIMRW.Hasegawa} (SST-CP).
By this replacement, the 1.85-m telescope was ready to use as a base of VLBI observations.

\subsubsection{Season 2016--2019}

\begin{figure} [tb]
\begin{center}
\begin{tabular}{c} %% tabular useful for creating an array of images 
\includegraphics[width=15cm, bb=0 0 883 452]{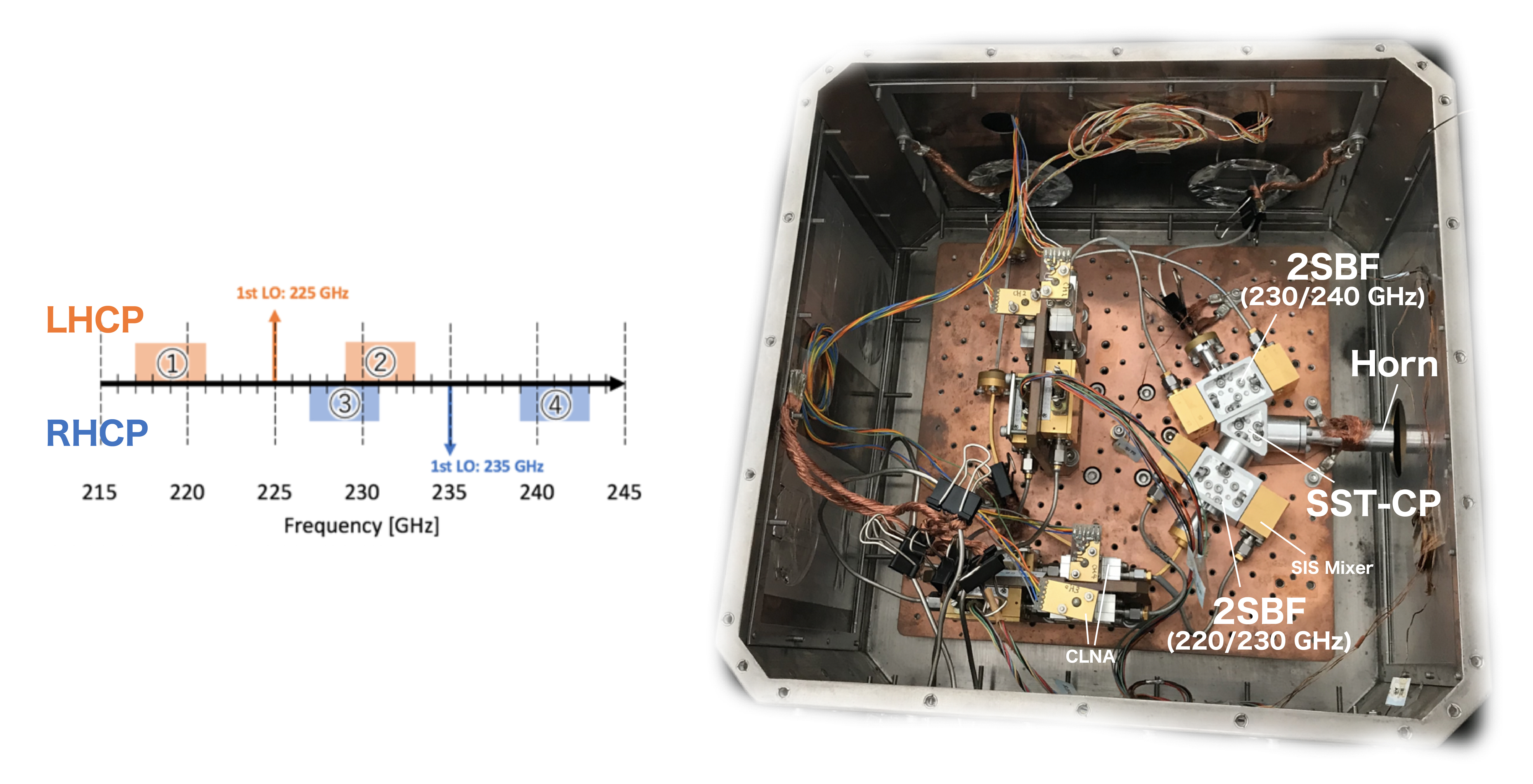}
\end{tabular}
\end{center}
\caption[example] 
%>>>> use \label inside caption to get Fig. number with \ref{}
{ \label{fig:rx2016} 
(Left) Schematic diagram of the frequency separation in the 220/230/240 receiver.
(Right) Photograph of the 220/230/240 receiver.
}
\end{figure}

By the season of 2015, the survey observations of CO $J=$2--1 was almost completed for the Galactic plane on the northern hemisphere (see also Section \ref{sec:co}), so we changed the science goal to survey weaker emission line in the 220 GHz band such as SO, CS, and CH$_{3}$OH.
In order to achieve the new theme, we developed the new filter system based on 2SBF (220/230/240 Rx).
In the system, the polarization of the received radio wave was firstly separated by the SST-CP, then signals of each of polarization are derived to the two 2SBFs which were tuned for the different RF frequency (Fig. \ref{fig:rx2016}): one for 220 and 230 GHz, and another for 230 and 240 GHz.
The typical system noise temperature ($T_{\rm sys}$) of the 220/230/240 Rx was measured to be 80, 90, 100, and 150 K for 220, 230 (LHCP), 230 (RHCP), and 240 GHz band, respectively, measured with the 1.85-m telescope at NRO.

\subsection{Telescope Control System}

The telescope and various equipment including the spectrometers are controlled and monitored on a Linux PC system via TCP/UDP/IP socket connections.
The whole system consists of the following components: 1, Hardware driver layer consisting of programs for controlling each of the instrument (e.g., telescope drives, encoders, receiver equipment); 2, Communication layer consisting of programs which command the instruments in cooperation to realize the observations; 3, Database layer consisting of programs that collect and store the data including the spectrometer output, environmental parameters, and operation log.

\subsubsection{Season 2009--2017}

\begin{figure} [b]
\begin{center}
\begin{tabular}{c} %% tabular useful for creating an array of images 
\includegraphics[width=15cm, bb=0 0 1015 189]{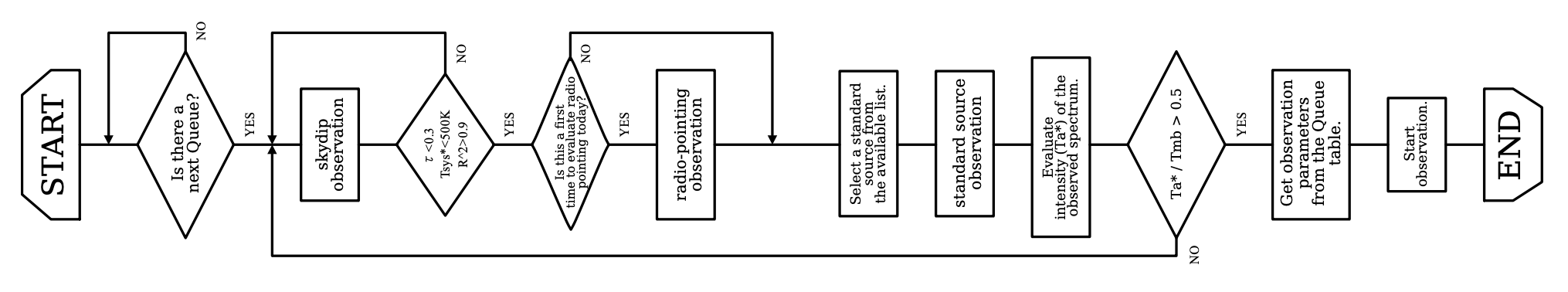}
\end{tabular}
\end{center}
\caption[example] 
%>>>> use \label inside caption to get Fig. number with \ref{}
{ \label{fig:queue} 
Flowchart of the queue based semi-automatic observation implemented in the observation system on the 1.85-m telescope.
}
\end{figure} 

\begin{figure} [tb]
\begin{center}
\begin{tabular}{c} %% tabular useful for creating an array of images 
\includegraphics[width=12cm, bb=0 0 869 531]{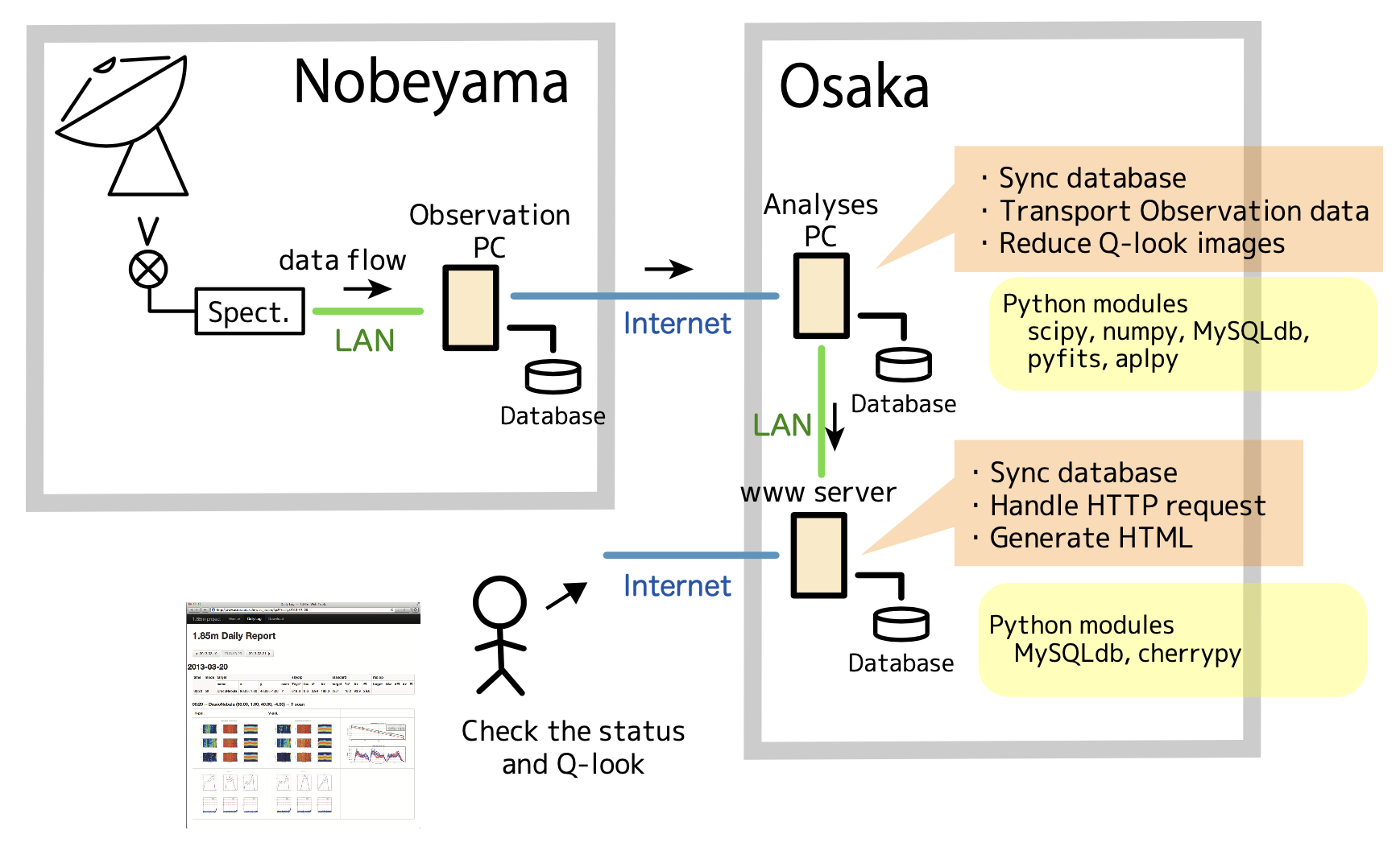}
\end{tabular}
\end{center}
\caption[example] 
%>>>> use \label inside caption to get Fig. number with \ref{}
{ \label{fig:pipeline} 
Schematic diagram of the web-based Q-look system.
}
\end{figure} 

The control system for the 1.85-m telescope was started with a following composition.
The hardware driver layer was mostly written in C language to control the PCI boards, and written as a TCP server.
The communication layer was written in Python, and the database was implemented integrally on MySQL database system.
The detail of the implementation of control system is described in Ref. \citenum{2013PASJ...65...78O}.
Based on the framework, the queue based semi-automatic operation was realized which reduced burden for the telescope operator and the time loss due to the careless mistake, and hence improved the actual observation efficiency (Fig. \ref{fig:queue}).
Furthermore, we implemented the automatic data analyzing pipeline that provides observation results graphically via the web interface soon after the finishing each observation (Fig. \ref{fig:pipeline}).

\subsubsection{Season 2018--2019}

From the observation season of 2018, we started to develop a new control system based on ROS\cite{2009.ROS} (robot operating system).
The ROS is one of the most prevailing open source framework to control robots and provides a very useful communication module (called Topic) which realizes many-to-many communications between processes and/or computers.
The reliability and scalability on the communication layer was improved by introducing the Topic based system.
In addition, aiming to realize further efficiency on developing time, we developed the fully python based hardware drivers, {\tt pypci}\footnote{https://github.com/ars096/pypci} and {\tt pyinterface}\footnote{https://github.com/ogawa-ros/pyinterface}.
The detail of the ROS based control system NECST is described in Ref. \citenum{2020SPIE.kondo}, and the performance evaluations of the new control system is ongoing.

\begin{figure} [b]
\begin{center}
\begin{tabular}{c} %% tabular useful for creating an array of images 
\includegraphics[width=15cm, bb=0 0 720 288]{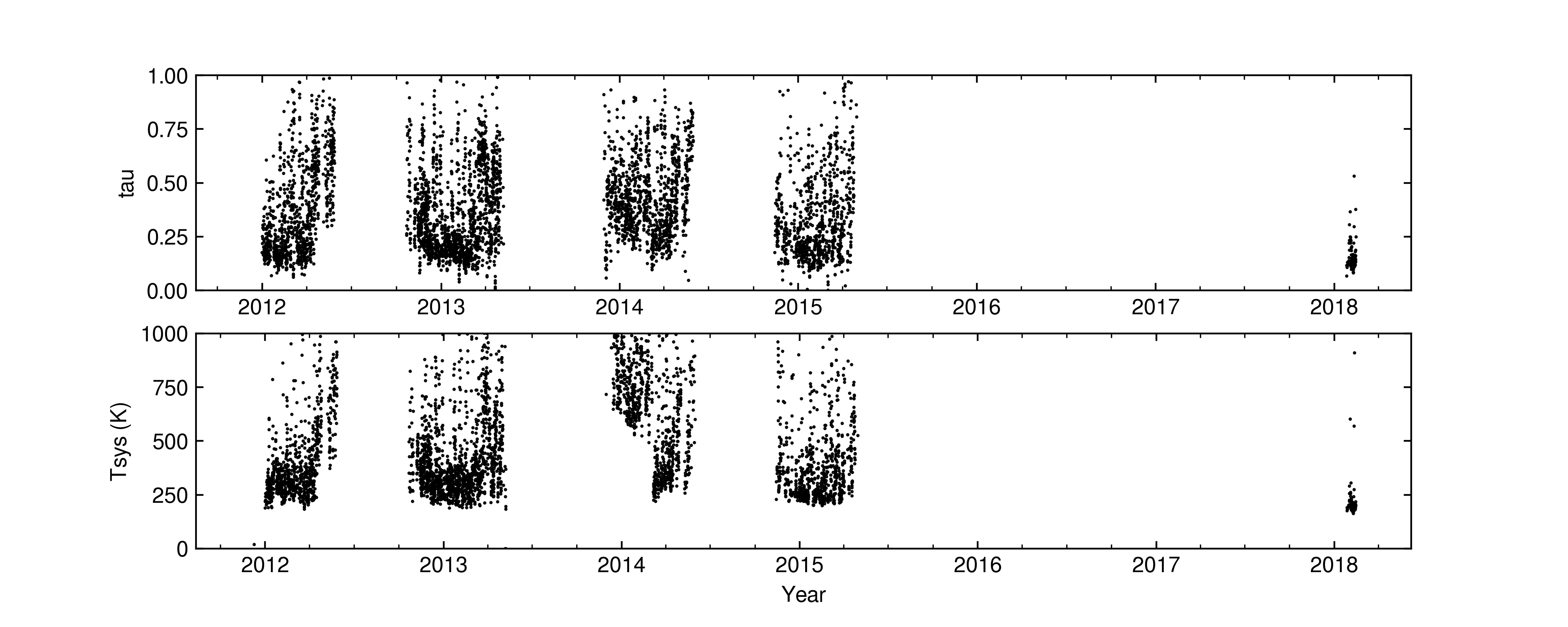}
\end{tabular}
\end{center}
\caption[example] 
%>>>> use \label inside caption to get Fig. number with \ref{}
{ \label{fig:tau} 
Distribution of the optical depth (upper) and the system noise temperature ($T_{\rm sys}$) including the atmosphere toward the zenith (lower) measured by the 1.85-m telescope with a frequency band of 230 GHz.
Before 2012, the integrated database system was not installed, so the environmental data before that time was lost.
}
\end{figure} 

\begin{figure} [H]
\begin{center}
\begin{tabular}{c} %% tabular useful for creating an array of images 
\includegraphics[width=14.5cm, bb=0 0 1217 1651]{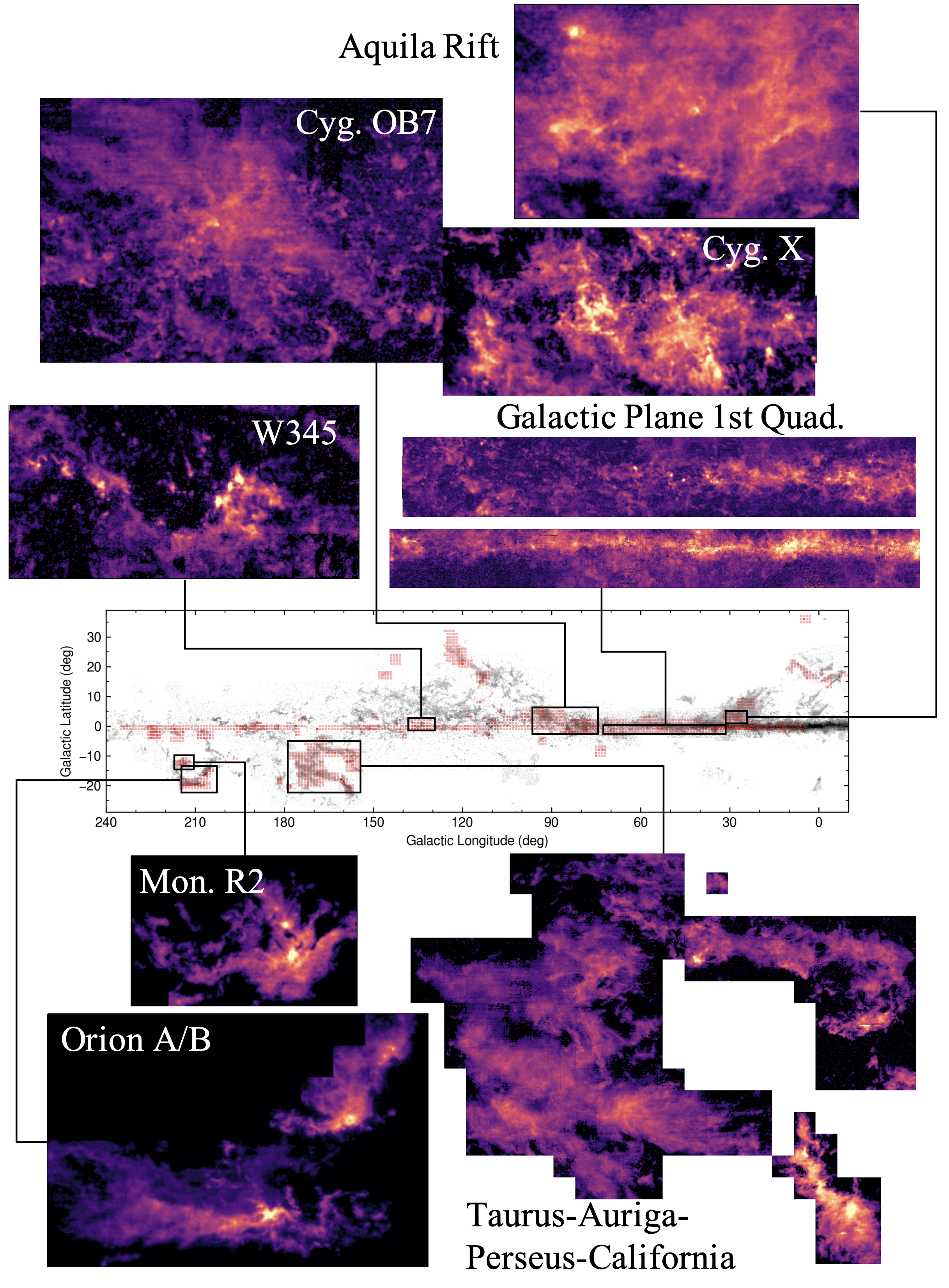}
\end{tabular}
\end{center}
\caption[example] 
%>>>> use \label inside caption to get Fig. number with \ref{}
{ \label{fig:co} 
Middle: Observation coverage of the 1.85-m CO $J=$2--1 survey. 
Background gray-scale image shows the distribution of $^{12}$CO $J=$1--0 taken by Ref. \citenum{2001ApJ...547..792D} as reference.
Red squares indicate observed area by the 1.85-m telescope.
Upper and lower: Some samples of the distributions of CO $J=$2--1 toward the famous star formation regions and the Galactic plane.
Color ranges are different each other.
}
\end{figure} 

\section{OBSERVATIONS}
\label{sec:obs}

The science operation was started on 2011.
Usually, the science observations were conducted during the winter season (from November to next April), and the telescope was shut-down during the summer season because the high humidity condition reduces observation efficiency from the ground.
Figure \ref{fig:tau} shows the distribution of the optical depth and the system noise temperature ($T_{\rm sys}$) including the atmosphere measured toward the zenith with a frequency band of 230 GHz at NRO.
About the 60\% of the observation seasons show a better sky condition than the threshold of $T_{\rm sys} < 400$ K and the optical depth of $<0.4$. 
For the seasons of 2010--2014 (Nov. 2010 to Apr. 2015), we performed the large scale survey of CO $J=$2--1, for the seasons of 2015--2016 were devoted to the demonstration experiments for the new instruments described in Section \ref{sec:rx}, and we conducted the observations of multiple line mapping in the season 2017 (Nov. 2017 to Apr. 2018).

\subsection{Survey Observations of CO J=2--1 Lines}
\label{sec:co}

Fig. \ref{fig:co} shows the survey coverage of the 1.85-m observations and some examples of obtained CO $J=$2--1 maps.
Totally 1800 deg$^{2}$ area (corresponding to 4.4\% of all-sky) was observed in the seasons of 2010, 2011, 2012, 2013, and 2014, and the observed areas for each season were 323, 160, 729, 303, and 307 deg$^{2}$, respectively.
All observations were conducted by the On-the-Fly (OTF) mapping mode with an angular spacing of 1$'$ grid.
The number of obtained spectra is 6,480,000.
The spatial resolution after the data reduction was $\sim3'$, and the typical noise level was $T_{\rm rms} \sim 0.8$ K with a velocity resolution of $\sim$0.1 km s$^{-1}$.
Using the CO data, studies on the star formation\cite{2013ApJ...768...72S, 2015ApJS..216...18N, 2016ApJ...818...59D, 2016A&A...589A..80H, 2016A&A...590A...2S, 2017ApJ...837..154N, 2017ApJ...845..105D, 2017A&A...608A..21S, 2017ApJ...844..138K, 2018AJ....156...84K}, and on the ISM\cite{2018A&A...609A.127S, 2019A&A...628A..44F} were promoted.

From the series of CO $J=$2--1 observations toward the GMCs in the Galaxy, we found that $^{12}$CO $J=$2--1 is easily becoming optically thick at the inner part of the GMC where the column density is higher than $\sim 3 \times 10^{21}$ cm$^{-2}$, and hence the ratio of $^{12}$CO $J=$2--1/$^{12}$CO $J=$1--0, $R^{12}_{2-1/1-0}$, is observed as $\sim 0.7$ which is a value expected to be observed in the condition of the local thermodynamic equilibrium (LTE) by the numerical simulations\cite{2017MNRAS.465.2277P}.
On the other hand, we found that the ratio of $^{13}$CO $J=$2--1/$^{13}$CO $J=$1--0, $R^{13}_{2-1/1-0}$, is a good tracer of the volume density even toward the inner region of the GMC.
For the case of Orion GMCs, the observed $R^{13}_{2-1/1-0}$ is in the range of 0.2--2.0, and its variation is clearly seen even toward the most inner part of the GMC where the column density is $\sim 10^{22}$ cm$^{-2}$\cite{2015ApJS..216...18N}.
The volume density of 800--2000 cm$^{-3}$ is estimated in the region including the inner part of the GMC by using LVG analysis.

\subsection{Spectral Line Surveys for 230 GHz Band}

\begin{figure} [b]
\begin{center}
\begin{tabular}{c} %% tabular useful for creating an array of images 
\includegraphics[width=15cm, bb=0 0 580 286]{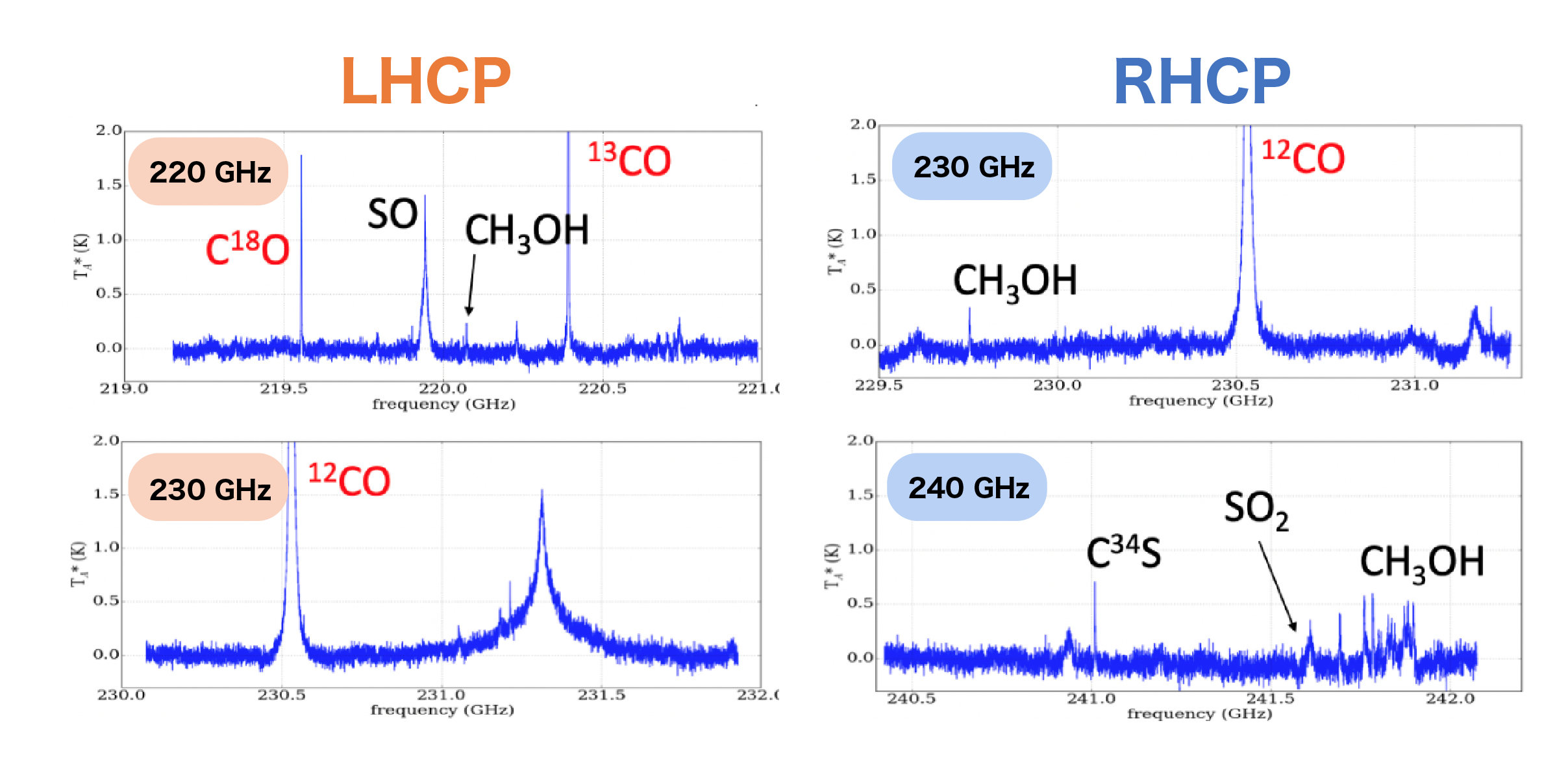}
\end{tabular}
\end{center}
\caption[example] 
%>>>> use \label inside caption to get Fig. number with \ref{}
{ \label{fig:line2016} 
Spectra observed toward the Orion-KL region.
}
\end{figure}

\begin{figure} [t]
\begin{center}
\begin{tabular}{c} %% tabular useful for creating an array of images 
\includegraphics[width=15cm, bb=0 0 593 357]{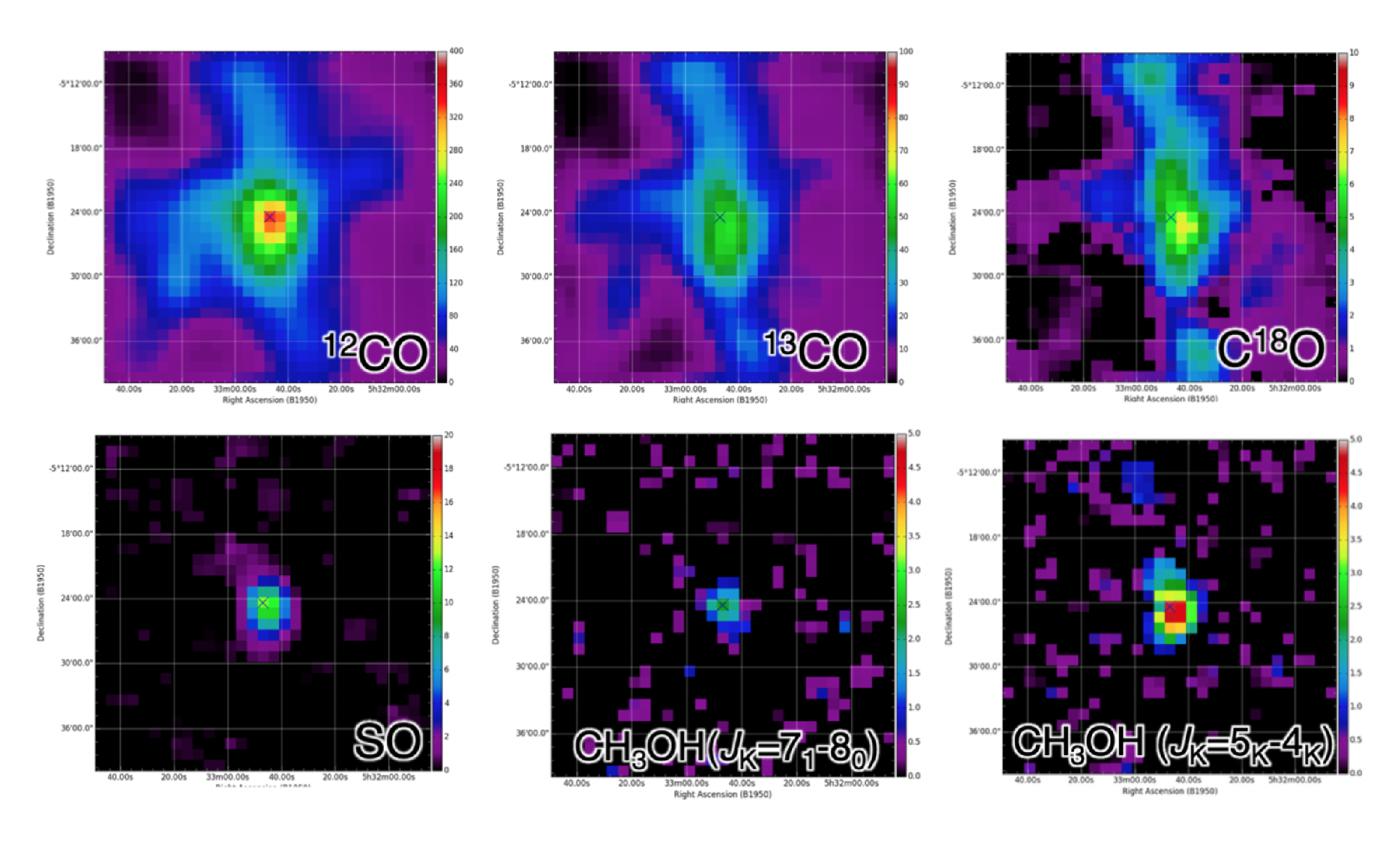}
\end{tabular}
\end{center}
\caption[example] 
%>>>> use \label inside caption to get Fig. number with \ref{}
{ \label{fig:map2016} 
Maps of $^{12}$CO, $^{13}$CO, C$^{18}$O, SO, and CH$_{3}$OH observed toward the Ori-KL region.
}
\end{figure}

In the early 2018, the test observations of the 2SBF receiver was conducted.
The receiver has four windows for observations: 219--221 GHz band including $^{13}$CO, C$^{18}$O, SO, and CH$_{3}$OH lines; 229.5--231.5 GHz band of the RHCP side including $^{12}$CO and CH$_{3}$OH lines; 230--232 GHz band of the LHCP side including $^{12}$CO line; and 240.5--242.5 GHz band including C$^{34}$S, SO$_{2}$, and CH$_{3}$OH lines.
Fig. \ref{fig:line2016} shows the spectra observed toward the Ori-KL region.
All spectra listed above were simultaneously detected.
Fig. \ref{fig:map2016} shows maps of $^{12}$CO, $^{13}$CO, C$^{18}$O, SO, and CH$_{3}$OH observed toward the Ori-KL region.
These results demonstrate the capability for the wide-band observations of the 2SBF receiver.

\subsection{Technology Verification on VLBI Observations in 230 GHz Band}

The experiment on VLBI observations in 230 GHz band was conducted on April 2015 between two bases using the 1.85-m telescope and the SPART telescope which is a single dish operation mode of Nobeyama millimeter array.
The length of the baseline was 150 m, and the instruments for the VLBI observations (e.g., a frequency standard generator, samplers) were brought to the telescope site in temporarily.
By observing the moon edge, VLBI fringe was successfully detected by the software correlator.
This success of the experiment led to the realization of the next VLBI experiment among the SPART 10-m, SRAO 6-m in Korea, and GLT 12-m in Greenland with a frequency of 230 GHz.

\section{FUTURE PLAN}
\label{sec:future}

\begin{figure} [t]
\begin{center}
\begin{tabular}{c} %% tabular useful for creating an array of images 
\includegraphics[width=15cm, bb=0 0 500 193]{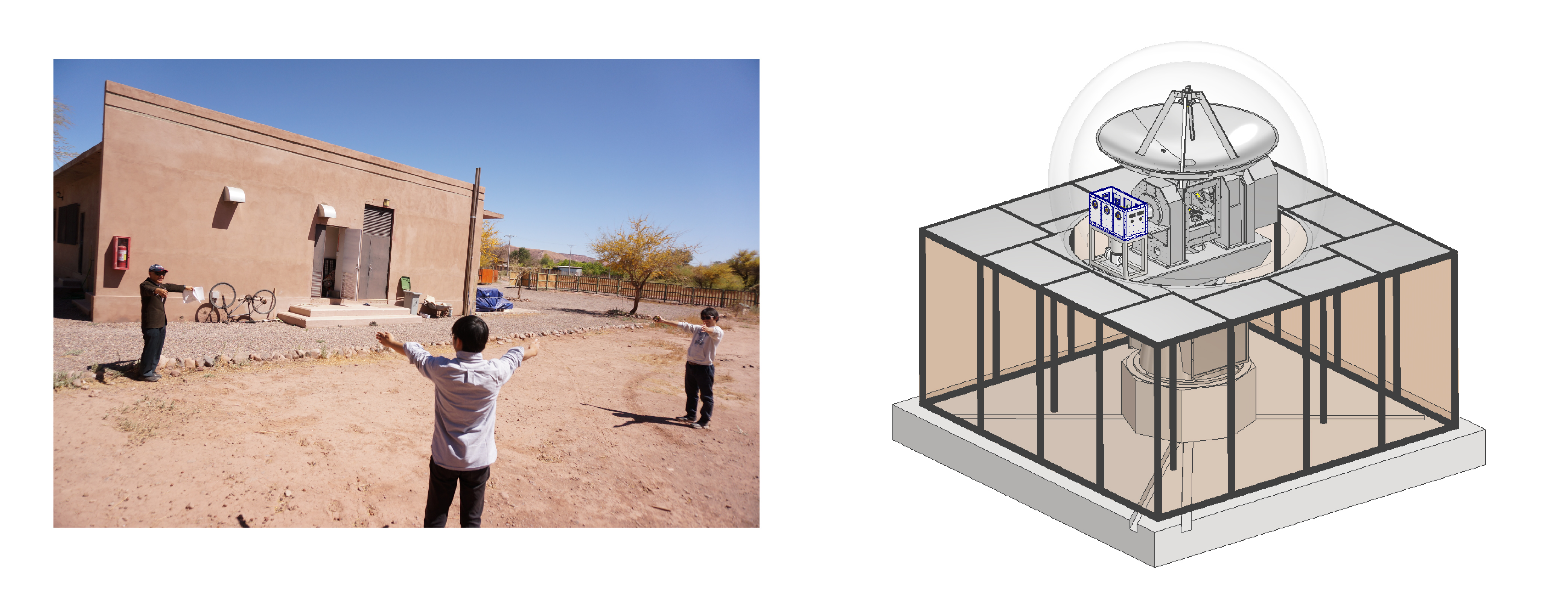}
\end{tabular}
\end{center}
\caption[example] 
%>>>> use \label inside caption to get Fig. number with \ref{}
{ \label{fig:tao} 
Left: Photograph of the planned location of the 1.85-m telescope installation in the base facility of the TAO project.
Persons, Prof. Hideo Ogawa, Mr. Yasumasa Yamasaki, and Mr. Takeru Matsumoto from left to right, indicate the size of the telescope base shown in the right figure.
Right: CAD drawing of the model of telescope base which will be constructed at the telescope site in SPdA.
}
\end{figure}

In order to extend the observation frequency band further wide, and to achieve simultaneous observations of 230 and 345 GHz band by sharing same single beam\cite{2020SPIE.masui, 2020SPIE.yamasaki}, we are planing to relocate the 1.85-m telescope to San Pedro de Atacama (SPdA), Chile (alt. 2400 m).
Although, the altitude of SPdA is not high as the Atacama high site where NANTEN2 and ASTE are installed, its weather condition would be almost sufficient for observations of 230 and 345 GHz band.
On the contrary, SPdA would be better choice in terms of the cost of construction and running.
The 1.85-m telescope will be installed in the base facility of the TAO project as part of the cooperative research with the university of Tokyo.
Fig. \ref{fig:tao} shows a photograph of the planned site of the 1.85-m telescope installation and a CAD drawing of the new telescope base which will be constructed to the site.
By this relocation, we will be able to perform the verification tests for the higher frequency instruments and to observe the higher frequency bands, as well as to observe southern sky including the Large/Small Magellanic Clouds.

\section{SUMMARY}
\label{sec:summary}

\begin{figure} [H]
\begin{center}
\begin{tabular}{c} %% tabular useful for creating an array of images 
\includegraphics[width=15cm, bb=0 0 792 216]{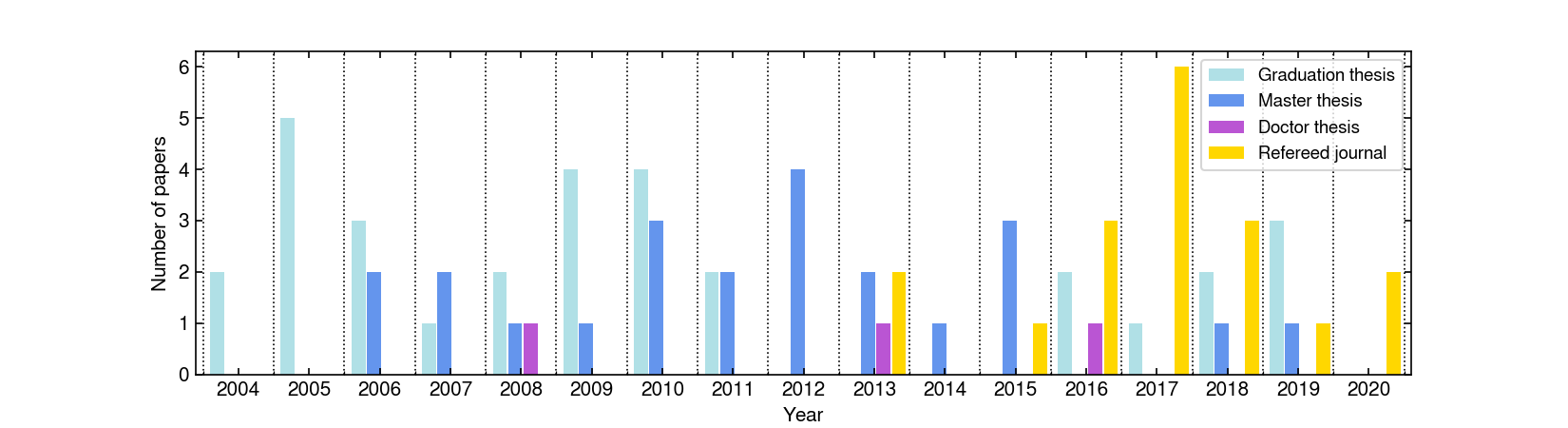}
\end{tabular}
\end{center}
\caption[example] 
%>>>> use \label inside caption to get Fig. number with \ref{}
{ \label{fig:pub} 
Bar chart indicating numbers of publications related to the 1.85-m telescope project.
}
\end{figure}

In the paper, we reviewed and summarized the activities of developments and observations on the 1.85-m telescope installed at Nobeyama Radio Observatory (NRO), Japan.
The construction of the telescope at NRO was started on 2007. 
The 2SB receiver with a RF of 230 GHz was installed, and its first light was achieved on September 2009 by observing the $^{12}$CO, $^{13}$CO, and C$^{18}$O $J=$2--1 lines simultaneously.
The science observation operations were started on November 2011, and the large-scale mapping observations of the $^{12}$CO, $^{13}$CO, and C$^{18}$O $J=$2--1 lines were continued to the observation season of 2014.
Finally, an area of 1800 deg$^{2}$ was observed and a number of 6,480,000 spectra was obtained.
During the period of science operations, the developments of the receivers and control systems were also continued.
Those developed systems were installed to the telescope in the summer season, when the science observations were stopped, and were tested for verification, and then, were used for scientific observations.
To date, 18 of refereed journal papers, 3 of doctor theses, 23 of master theses, and 31 of graduation theses were published related to the telescope or by using the data obtained as the CO survey (Fig. \ref{fig:pub}).

\acknowledgments % equivalent to \section*{ACKNOWLEDGMENTS}     

The 1.85-m telescope project is promoted on a lot of contributions of people engaged in this project, so the authors would like to thank all those people, Takafumi Kojima, Taku Nakajima, Toshihisa Tsutsumi, Ryoko Amari, Kozuki Yuto, Noriaki Arima, Ryosuke Kiridoshi, Takao Matsumoto, Shigeki Osaki, Touga Shiori, Minato Kozu, Yuya Ota, Yoshiharu Kojima, Akio Hashizume, Akihito Minami, Tsubasa Sakaguchi, Hidetoshi Tsuji, Tetsuya Katase, Shimpei Yashima, Masato Kunizane, Masahiro Minowa, Yukimasa Takenaka, Shinpei Yashima, Hirotaka Kurimoto, Atsushi Ezaki, Kazutoshi Maruyama, Hirofumi Okuno, Takashi Nohara, Kiyoko Tsuji, Yoshihide Tokko, Kazuki Toki, Tatsuro Nakayama, and Jun Korogi.
We are also grateful to Shigeru Fuji, Akira Mori and Hiroyuki Iwashita, and the entire staff of the Nobeyama Radio Observatory for their useful support. 
This work was supported by JSPS KAKENHI Grant Numbers JP18H05440, JP15K05025, JP26247026, JP14J12320, JP22244014, and JP15071205, by the Mitsubishi Foundation and the Toray Science Foundation.

% References
\bibliography{main} % bibliography data in report.bib
\bibliographystyle{spiebib} % makes bibtex use spiebib.bst

\end{document}